# Kinematical properties of the thick disk of the Galaxy

By Devendra Ojha

Observatoire de Besançon, BP 1615, F-25010, France †

We try here to find new constraints on thick disk population using samples at intermediate latitude and North Galactic Pole, which include photometry and proper motions. The algorithm SEM (Stochastic-Estimation-Maximization; Celeux & Deibolt 1986) was used to deconvolve the stellar components up to large distances above the plane, allowing to study of their statistical properties independently. The multivariate discriminant analysis (MDA) is used to qualify the thick disk using observations in multidimensional space (V,B-V,U-B, $\mu_l$ & $\mu_b$).

## 1. Samples

The data sets from the following 6 surveys have been used to derive the kinematical and structural parameters of the thick disk population :

- field in the direction of galactic anticentre ($l = 167°$, $b = 47°$ [Ojha et al. 1994a]); hereafter GAC.

- field in the direction of galactic centre ($l = 3°$, $b = 47°$ [Ojha et al. 1994b]); hereafter GC.

- field in the direction of galactic antirotation ($l = 278°$, $b = 47°$ [Ojha et al. 1994c]); hereafter ELLIPSOID.

- field in the direction of NGP-I ($l = 58°$, $b = 80°$ [Soubiran 1993]).

- field in the direction of NGP-II ($l = 42°$, $b = 79°$ [Kharchenko et al. 1994]).

- field in the direction of NGP-III ($l = 124°$, $b = 87°$ [Kharchenko et al. 1994]).

## 2. Methods

### 2.1. SEM

The aim of SEM algorithm is to resolve the finite mixture density estimation problem under the maximum likelihood approach using a probabilistic teacher step. Through SEM one can obtain the number of components of the gaussian mixture, its mean values, dispersions and the percentage of each component with respect to the whole sample. Applying the SEM algorithm on the real data sets from 6 *in situ* surveys, a thick disk population has been identified in different distance bins (up to z~3.5 kpc). Figure 1 shows the asymmetric drift measurements of thick disk population from selected studies as a function of z-distances. We find a similar value of the asymmetric drift of thick disk from the NGP and intermediate latitude data sets. As can be seen in figure 1, no clear dependence with z-distance is found in the asymmetric drift measurements of thick disk population. However a gradient can be seen in V-velocity if *no separation* was made between the 3 populations (thin disk, thick disk and halo). Vertical bars indicate the error $\frac{\sigma_V}{\sqrt{N}}$ in V velocity. Where N is the number of stars in each distance bin.

† Contributed talk presented at the workshop of The Formation of the Milky Way, Granada (Spain) 5-10 September 1994.

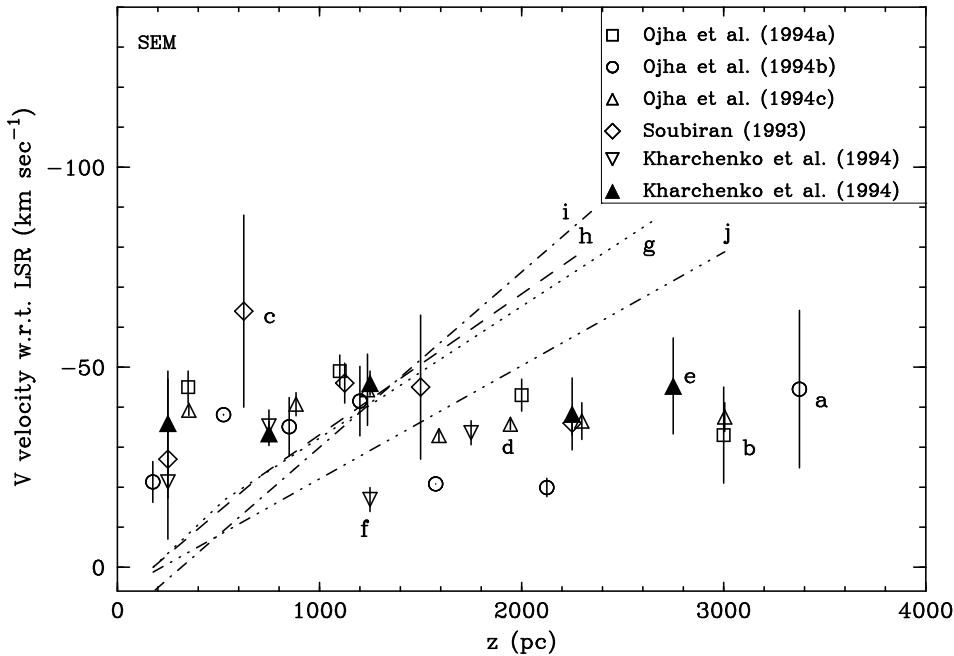

FIGURE 1. The measured asymmetric drift of thick disk population plotted as a function of z distances from the proper motion selected samples (open circle): a– Ojha et al. (1994b); (open square): b– Ojha et al. (1994a); (open diamond): c– Soubiran (1993); (open triangle up): d– Ojha et al. (1994c); (filled triangle up): e– Kharchenko et al. (1994); (open triangle down): f– Kharchenko et al. (1994). The dotted (g), dashed (h), dashed-dotted (i) and dashed-dotted-dotted (j) lines represent GC, GAC, NGP (Soubiran 1993) and ELLIPSOID fields, respectively with *no separation* between the three populations.

## 2.2. *MDA*

We have done the multivariate discriminant analysis using the two data sets and the Besançon model of population synthesis (Robin & Crézé 1986, Bienaymé et al. 1987). The observed data in 5-dimensional space (V,B-V,U-B, $\mu_l$, $\mu_b$) have been used for this analysis. The two data sets we used are the GC field ($l = 3°$, $b = 47°$) and GAC field ($l = 167°$, $b = 47°$). The U-B and $\mu_l$ parameters are necessary to make a good discrimination between the three populations, because U-B is sensitive to the metallicity and $\mu_l$ is parallel to the V velocity and discriminate the populations by their asymmetric drift.

We used the model simulations to find the best discriminant axes where to project the data in order to separate the thick disk population from the thin disk and halo. To avoid too large Poisson noise in the Monto Carlo simulations, we computed at least 10 simulations in 100 square degrees for each of the models tested in our analysis. The best discriminant axis for the circular velocity of 180 km sec$^{-1}$ of thick disk in the direction of GC is given by :

$x = 0.024(B - V) + 0.139(U - B) - 0.079V - 0.310\mu_l - 0.069\mu_b$

To quantitatively estimate the adequacy of the models with various circular velocities, we applied a $\chi^2$ test to compare the distribution of the sample on the discriminant axis with a set of model predicted distribution assuming different circular velocities of thick disk. Table 1 shows the values of the probability (in sigmas) of each model to come from the same distribution as the observed sample. Where sigma is an approximation to the distribution of $\chi^2$ (Kendall & Stuart 1969). The most probable value for the circular velocity of thick disk comes out to be 180 km sec$^{-1}$, corresponding to a lag or asymmetric drift of thick disk of the order of 40±10 km sec$^{-1}$.

| $V_{thd}$ (km sec$^{-1}$) | Lag (km sec$^{-1}$) | $\chi^2$ (in sigmas) GC | GAC |
|---|---|---|---|
| 150 | 70 | 8.7 | 6.8 |
| 165 | 55 | 6.8 | 5.7 |
| 175 | 45 | 4.1 | 3.8 |
| 180 | 40 | 4.3 | 3.4 |
| 185 | 35 | 4.9 | 3.5 |
| 190 | 30 | 5.5 | 4.0 |
| 215 | 5 | 6.7 | 5.1 |

TABLE 1. $\chi^2$ test for models with different circular velocities of thick disk. $\chi^2$ is given in number of sigmas. Lag or asymmetric drift is w.r.t. the LSR, assuming $V_{LSR}$=220 km sec$^{-1}$.

The distribution over the discriminant axis of observed stars (solid points) towards GAC with best model predictions (with three populations) is shown in figure 2.

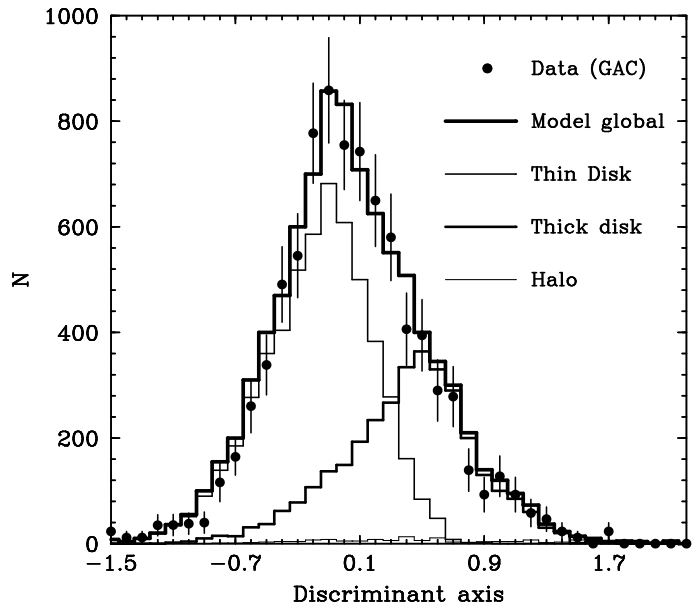

FIGURE 2. Distribution over the discriminant axis of observed stars (solid points) towards GAC direction; best model predictions (solid full thick line) and model predicted stars according to their populations (thick line : thin disk, full thick line : thick disk and thin line : halo)

## 3. Conclusion

We have deduced the new estimates of the kinematical parameters of thick disk population. The thick disk population is found to have a rotational velocity of $V_{rot} = 177$ km sec$^{-1}$, and with velocity dispersions $(\sigma_U, \sigma_V, \sigma_W) = (67, 51, 46)$ km sec$^{-1}$. Our data are consistent with no dependence of the thick disk's asymmetric drift with distance above the galactic plane (up to z = 3 kpc). We obtain a unique value for the asymmetric drift of thick disk in two directions (galactic centre and anticentre), showing that no radial gradient seems to occur on the asymmetric drift of thick disk on a base of 3 kpc around the Sun. From the number ratio of the thick disk stars in a pair of direction (towards galactic centre and anticentre), we deduce the scale length $h_R$ of the thick disk, which is found to be 3.6±0.5 kpc. The density laws for stars with $3.5 \leq M_V \leq 5$ as a function of distance above the plane, follow a single exponential with scale height of $\sim 260$ pc for $150 \leq z \leq 1200$ pc, and a second exponential with scale height $\sim 770$ pc for z distances from $\sim 1200$ pc to at least 3000 pc. We identify the 260 pc scale height component as a thin disk, and the 770 pc scale height component as a thick disk (Ojha et al. 1994d).

This research work was partially supported by the Indo-French Center for the Promotion of Advanced Research (IFCPAR), New-Delhi (India). We thank Dr. E. Schilbach for letting us use their data.

**Carney :** Since your kinematics depend solely on tangential velocities, your absolute magnitudes are very important. Did you use a metallicity sensitivity ? if not, how might that affect your results ?

**Ojha :** Since the local value of the vertical metallicity gradient is not well determined, we assumed a model of vertical metallicity gradient as in Kuijken & Gilmore (1989). This model gives an indicative rather then conclusive estimates of the true gradient. The relation of the sensitivity of $M_V$ to [Fe/H] was taken from Laird et al. (1988) and the UV excess- metallicity ralation was taken from Carney (1979). In our distance determination, we did not consider the radial metallicity gradient. The overall error is about 10-20% in the velocity, which is mainly due to the photometric uncertainties.

**Majewski :** Do you allow for kinematical gradients in your second analysis (the multivariate discriminant analysis not the SEM) or do you assume discrete components only?

**Ojha :** In case of MDA, we have selected a subsample of stars (where the majority of thick disk stars exits) with the help of Besançon model. So we assume the thick disk as a discrete component and did not allow the kinematical gradients in this analysis.

**Carney :** Did you try to derive the velocity ellipsoïd for the thin disk ? Your thick disk velocity ellipsoïd suggests near-equality of the $\theta$ and $z$ components, which is different than the thin disk results. This is important since it bears on the continuity or discontinuity of the thin and thick disks.

**Ojha :** In case of the thin disk we did not derive the $z$ component. We found that in the nearer distance bins, the gaussian fit (using SEM) shows that the thin disk-like population was itself devided into two components. However, we derived the $\pi$ and $\theta$ components for the thin disk which are found to be : $\sigma_\pi$=40 km/s, $\sigma_\theta$=30 km/s.